\documentclass[aps,pra,twocolumn,amsfonts,amssymb,amsmath,showpacs,
floatfix,nofootinbib,groupedaddress,superscriptaddress,citesort]{revtex4}
\usepackage{mathrsfs}
\usepackage{amsfonts}
\usepackage{amstext}
\usepackage{amsmath}
\usepackage{amssymb}\usepackage{subfigure}
\usepackage{bm}% bold math
\usepackage{bbm}\usepackage{threeparttable}

\usepackage{graphicx}
\def\qed{\leavevmode\unskip\penalty9999 \hbox{}\nobreak\hfill
     \quad\hbox{\leavevmode  \hbox to.77778em{%
              \hfil\vrule   \vbox to.675em%
               {\hrule width.6em\vfil\hrule}\vrule\hfil}}
     \par\vskip3pt}

\def\ra{\rangle}
\def\la{\langle}
\def\bb{\mathbb}

\begin{document}

%\preprint{APS/123-QED}
\title{Quantifying imaginarity in terms of pure-state imaginarity}% Force line breaks with \\

\author{Shuanping Du}
\email{dushuanping@xmu.edu.cn} \affiliation{School of Mathematical
Sciences, Xiamen University, Xiamen, Fujian, 361000, China}

\author{Zhaofang Bai}\thanks{Corresponding author}
\email{baizhaofang@xmu.edu.cn} \affiliation{School of Mathematical
Sciences, Xiamen University, Xiamen, Fujian, 361000, China}

\begin{abstract}

Complex numbers are widely used in quantum physics and are indispensable components
for describing quantum systems and their dynamical behavior. The resource theory of imaginarity
has been built recently, enabling a systematic research of complex numbers in  quantum
information theory. In this work, we develop two theoretical methods for quantifying imaginarity, motivated
by recent progress within resource theories of entanglement and coherence. We provide quantifiers of imaginarity by the convex roof construction and  quantifiers of the imaginarity  by the
least imaginarity of the input pure states under real operations. We also apply these tools to study the state
conversion problem in resource theory of imaginarity.

%\begin{description}
%\item[PACS numbers]03.65.Ud, 03.67.-a, 03.65.Ta.
%\end{description}
\end{abstract}

\maketitle
%\end{CJK*}
%\tableofcontents

\section{Introduction and main results}  Quantum mechanics is traditionally formulated using complex numbers.
Hickey and Gour \cite{Hickey1} introduced the imaginarity of a quantum state
as a quantum resource theory \cite{Hickey2}. In spite of this rigorous resource
theoretic treatment, its usefulness is initially unclear.
Indeed, there has been a long-standing debate since the inception of quantum mechanics
about why quantum mechanics uses complex numbers rather than solely real
numbers. Some early results in the literature suggest that imaginarity
has no relevant role, as, for instance, various real-only formulations
of quantum mechanics were known \cite{Stueckelberg1,Stueckelberg2,Wootters1,Wootters2,Aleksandrova,Wootters3}.

Surprisingly, Renou et al. recently have
found an experimental scenario that cannot be exactly modeled using
quantum theory with real-valued amplitudes only \cite{Renou}.
These findings in relevant scenarios have been experimentally verified and proposed \cite{Fan,Pan1,Pan2,Bednorz,Fan2}.
From then on, there is a burst of investigations on the imaginarity resource theory.
It is shown that imaginarity as an important resource have crucial
 effects in certain discrimination tasks \cite{Guo1},
hiding and masking \cite{Zhu}, multiparameter metrology \cite{Miyazaki}, machine learning \cite{Sajjan}, pseudorandomness \cite{Haug},
outcome statistics of linear-optical experiments \cite{Jones},
Kirkwood-Dirac quasiprobability distributions \cite{Budiyono1,Budiyono2,Budiyono3,Wagner}
, weak-value theory \cite{Wagner2} and nonlocal advantages of quantum imaginarity \cite{Fei}.

A resource theory is fundamentally determined by  two
basic elements: free states and free operations \cite{Hickey2,Adesso}. The free states in resource theory of imaginarity  are {\it real states},
i.e., quantum states with a real density matrix $\langle m |\rho |n\rangle\in{\mathbb R}$
for a fixed reference basis $\{|m\ra\}_{m=1}^d$ of $\mathcal H$. The free operations are {\it real operations},
i.e., quantum operations $\Phi(\rho) =\sum_j K_j\rho K_j^{\dag}$ with real
Kraus operators: $\langle m|K_j|n\rangle\in{\mathbb R}$.

One central question in resource theory of imaginarity is quantitative analysis that captures the resource character of physical traits in
a mathematically rigorous fashion. A nonnegative function ${\mathcal I}$ on quantum states is
called an imaginarity measure if ${\mathcal I}$ satisfies the following
conditions (I1) to (I4) \cite{Hickey1,Guo2,Guo1}.

\noindent(I1) Non-negativity: $\mathcal I (\rho) \geq 0$, and $\mathcal I (\rho) = 0$ for any real state $\rho$.
	
	\noindent(I2) Monotonicity: $\mathcal I (\Phi(\rho))\leq  \mathcal I (\rho)$ whenever $\Phi$ is a real operation.
	
		\noindent(I3) Probabilistic monotonicity: $\sum_j p_j{\mathcal
		I}(\rho_j)\leq {\mathcal I}(\rho),$ where $p_j=tr (K_j\rho
	K_j^{\dag})$, $\rho_j=\frac {1}{p_j}K_j\rho K_j^{\dag}$, for all
	$\{K_j\}$ with $\sum_j K_j^{\dag}K_j=I$ and $K_js$ are real operators.
	
	\noindent(I4) Convexity: ${\mathcal I}(\sum_jp_j\rho_j)\leq \sum_jp_j{\mathcal
		I}(\rho_j)$ for any ensemble $\{p_j,\rho_j\}$.

Note that (I3) and (I4) together imply (I2). In Ref.
\cite{Li}, the authors considered the condition (I5) below.

\noindent (I5) Additivity for direct sum states:
$${\mathcal I}(p\rho_1 \bigoplus (1-p)\rho_2) = p{\mathcal I}(\rho_1)+(1-p){\mathcal I}(\rho_2). $$
It is shown that (I3) and (I4) are equivalent to
(I2) and (I5) \cite{Li}.  ${\mathcal I}$ will be called an
imaginarity monotone if it satisfies all the conditions but (I4),
which is similar to the entanglement monotone \cite{Hb1}.

 Several imaginarity measures have
been found, such as geometric imaginarity \cite{Guo2}, the robustness of imaginarity \cite{Guo2}, the trace norm and $l_1$ norm of imaginarity \cite{Hickey1,Guo1,Gao}, the relative entropy of imaginarity \cite{Li}, imaginarity based on the fidelity \cite{Guo2,Kondra,Xu} and Tsallis relative entropy \cite{Xu2}. It is obvious that different quantifications not
only provide different computability, but also imply different
operational meanings. However, as far as a finite number of imaginarity measures are considered, the quantification of imaginarity is still in early stages.
 How to explore the new understandings
of imaginarity remains still a significant and attractive topic in
the resource theory of imaginarity.

In this paper, we firstly provide quantifiers of imaginarity by the convex roof construction,
a traditional and effective method in the quantification of entanglement \cite{Uhlmann,Bennett,Vidal} and coherence \cite{Du1}.
Based on general properties of our recipe and their connection to the
state conversion problem, no-go theorem of
imaginarity convertibility on general mixed states are built.
Secondly, we consider that some pure states undergo real operations \cite{Guo1} and finally become the common objective state. It is
shown that the imaginarity of the objective state can be well described by the least imaginarity of the pure input states. So a imaginarity monotone is established by only effectively quantifying the input pure states. This automatically endows the imaginarity monotone with an operational meaning. Thirdly, the inner relation between convex roof quantifiers of imaginarity and imaginarity monotone induced by input pure states is revealed in detail.

In the first, let us treat convex roof quantifiers of imaginarity.
  Assume that 	$f:[0,1]\rightarrow [0,1]$ is a  function
	satisfying the following conditions:
		
	(i) $f(1)=0$;
	
	(ii) $f$ is decreasing;
	
	(iii) $f$ is concave: $f(\lambda {x}+(1-\lambda)
	{y})\geq \lambda f({x})+(1-\lambda)f({y})$
	for all $\lambda\in [0,1]$ and all $x,y\in [0,1]$.
	
	For any pure state $|\psi\rangle\langle\psi|$, define
	\begin{equation}
		{\mathcal I}_f(|\psi\rangle)=f(|\la \psi^*|\psi\ra|),\end{equation} here $|\psi^*\rangle$ is the complex conjugate of $|\psi\rangle$.
	For any mixed state $\rho$, define
	\begin{equation}
		{\mathcal I}_f(\rho)=\min _{p_i,|\psi_i\rangle}\{\sum_i
	p_i{\mathcal I}_f(|\psi_i\rangle):\ \rho=\sum\limits_i
	p_i|\psi_i\rangle\langle\psi_i|\}.
	\end{equation}
	
\textbf{ Theorem 1.} {\it For every function $f$ with {\rm(i-iii)}, the ${\mathcal I}_f$ defined in {\rm(1)} and {\rm(2)} is an
	imaginarity measure satisfying {\rm(I1)-(I5)}. Conversely,
the restriction  of any imaginarity measure satisfying
{\rm (I1), (I3), (I4))} to pure states is identical to ${\mathcal I}_f$ for some $f$.}

It is well-known that  both convex roof quantifiers of entanglement \cite{Uhlmann,Bennett,Vidal} and convex roof quantifiers of coherence \cite{Du1} are dependent on
real symmetric concave functions on probability distributions, while convex roof quantifiers of imaginarity are dependent on monotonically decreasing concave functions on $[0,1]$ from Theorem 1. This shows that convex roof quantifiers of imaginarity are more handy for operation. In addition, if $f$ is strictly decreasing, then ${\mathcal I}_f$ is faithful, that is, ${\mathcal I}_f (\rho) = 0$ if and only if $\rho$ is a real state.

Using Theorem 1, one can construct many interesting imaginarity measures. We provide two examples here.

Example 1: Let $f(x)=\frac{1-x}{2}$, then
\begin{equation}\mathcal I_f(|\psi\rangle)=\frac{1-|\langle\psi^*|\psi\rangle|}{2}\end{equation}
for a pure state $|\psi\rangle$. For mixed states, $\mathcal I_f$ is defined by convex roof method. Actually,
$\mathcal I_f$ equals to the key geometric imaginarity $\mathcal I_G(|\psi\rangle)=1-\max_{|\phi\rangle\in \mathcal R}|\la\phi|\psi\ra|^2$ \cite{Guo1,Guo2,Kondra}.

Example 2: Let $f(x)=\sqrt{\frac{1-x^2}{2}}$, assume $\rho=|\psi\rangle\langle\psi|=\sum_{m,n}\rho_{mn}|m\rangle\langle n|\  (\rho_{mn}=\langle m|\rho |n\rangle)$, a direct computation (see Proposition 3 in the appendix) shows
\begin{equation}\mathcal I_f(|\psi\rangle)=\sqrt{\sum_{m\neq n}({\rm Im}\rho_{mn})^2},\end{equation} here ${\rm Im}\rho_{mn}$ denotes the imaginary part of $\rho_{mn}$. This shows the restriction of $\mathcal I_f$
on pure  states is imaginarity measures induced by $l_2$ norm. Furthermore, it satisfies probabilistic monotonicity. An interesting observation in quantum coherence is that a desirable function induced by $l_2$  norms does not satisfy probabilistic monotonicity \cite{Baumgratz}.

%It has been shown $\mathcal I_f=\mathcal I_G$ if $f(x)=\frac{1-x}{2}$.
By Theorem 1, imaginarity measures for pure states correspond to monotonically decreasing concave functions on $[0,1]$.
The corresponding relationships between several well-known imaginarity measures and decreasing concave functions are revealed (see Table I), and  so we can induce these measures for pure states by convex roof construction. However,  quantifiers of imaginarity by the convex roof construction are indeed different from these known imaginarity measures on mixed states. Let us recall the notations and definitions of these measures.
The robustness of imaginarity $\mathcal I_{R}$ \cite{Guo2},
relative entropy of imaginarity \cite{Li}, denoted by $\mathcal I_r$,
$\mathcal I_F$ based on the fidelity \cite{Guo2,Kondra,Xu} and imaginarity measure
based on Tsallis relative entropy $\mathcal I_{T,\mu}$ \cite{Xu2}. They are defined as
\begin{equation}\mathcal I_R(\rho)=\min_{\tau}\{s\geq0:\frac{\rho+s\tau}{1+s}\in \mathcal R, \tau\in\mathcal S(H)\},\end{equation}
\begin{equation} \mathcal I_r(\rho)=S(\text{Re} \rho)-S(\rho),\end{equation}
\begin{equation} \mathcal I_F(\rho)=1-F(\rho,\rho^*),\end{equation}
\begin{equation}\mathcal I_{T,\mu}(\rho)=1-\text{Tr}[\rho^{\mu}(\rho^*)]^{1-\mu},\end{equation}
where $\text{Re} (\rho)$ is the real part of $\rho$,  $\rho^*$ denotes the complex conjugate of $\rho$, $\mu\in(0,1)$ and  $F(\rho,\sigma)=1-\text{Tr}\sqrt{\sqrt{\rho}
	\sigma\sqrt{\rho}}$ is the fidelity of $\rho$ and $\sigma$.
	
\begin{center}	
	\begin{table}[htbp]
		\centering
		\caption{ Corresponding relationship }
		\begin{threeparttable}\begin{tabular}{|c|c|c|c|c|c|}\hline
				imaginarity measure  &  $\mathcal I_G$ & $\mathcal I_R$ & $\mathcal I_r$ & $\mathcal I_F$ & $\mathcal I_{T,\mu}$ \\ \hline
				$ f(x)$ & $\frac {1-x}2$ & $\frac {\sqrt{1-x^2}}2 $ & $h(x)$ & $1-x $& $1-x^2 $ \\ \hline
			\end{tabular}
 {\small here $h(x)= -\frac{1+x}2\log (\frac{1+x}2)-\frac{1-x}2\log (\frac{1-x}2) $}.
\end{threeparttable} \end{table}\end{center}

The conversion of imaginary states is another important
research direction. Given two imaginary states $\rho$ and $\sigma$, one of the main questions studied within the resource theory of imaginarity is whether either $\rho$ can be transformed into $\sigma$ or vice
versa by real operations, i.e., identifying what
extra resources are necessary to enable such transformations \cite{Guo1,Guo2}. In fact, imaginarity quantification and  state convertibility are in fact closely connected.  By (I2), if a state $\rho$ can be converted into $\sigma$ via real operations, then \begin{equation}{\mathcal I}(\sigma)\leq  {\mathcal I}(\rho)\end{equation} for any imaginarity
monotone ${\mathcal I}$. On the other hand, the fact that  (9) holds for some imaginarity monotone ${\mathcal I}$ does not guarantee that the transformation $\rho \rightarrow\sigma$ is
possible via real operations. The aim of state convertibility is to find a
complete set of coherence monotones $\{{\mathcal I}_s\}$ which can completely classify states transformation, i.e.,
\begin{equation}\rho\rightarrow \sigma\Leftrightarrow {\mathcal I}_s(\rho)\geq {\mathcal I}_s(\sigma)\text{\ for all } s.\ \end{equation}

In \cite{Guo1}, the authors have answered the question of  pure-state convertibility
in terms of geometric imaginarity:  For pure states
$|\phi\rangle, |\psi\rangle$,
 \begin{equation}|\psi\rangle \xrightarrow{{\rm RO}}|\phi\rangle
\quad \text{iff} \quad {\mathcal I}_f(|\phi\rangle)\leq {\mathcal I}_f(|\psi\rangle),\end{equation}
 here RO denotes real operations and $f(x)=\frac{1-x}{2}$.
For one qubit case, mixed-state convertibility  has been considered in \cite{Guo2}, and the result can be restated in terms of imaginarity monotone as follows.
Let $${\mathcal I}_f(\rho)=|\text{Im}\rho_{12}|, \quad {\mathcal I}(\rho)=\left\{\begin{array}{cc}
             0, &\rho_{11}\rho_{22}=0\\

             \frac{|\text{Im}\rho_{12}|^2}{\rho_{11}\rho_{22}-|\text{Re}\rho_{12}|^2},& \rho_{11}\rho_{22}\neq 0,\end{array}\right.$$ where $f(x)=\frac{\sqrt{1-x^2}}{2}$.
It is shown that
\begin{equation}\rho \xrightarrow{{\rm
RO}}\sigma \quad \text{iff } {\mathcal I}_{f}(\rho)\geq {\mathcal I}_{f}(\sigma), {\mathcal I}(\rho)\geq {\mathcal I}(\sigma).\end{equation}

This motivates us to study mixed-state convertibility in multiple qubits case. From (11) and (12), a key question is that whether finite  number of measure conditions are sufficient to determine the existence of real operations between mixed
states? The following theorem implies that the answer is negative.
A parallel result in entanglement and coherence theory is the infinite number of measure conditions for  mixed-state conversion \cite{Gour,Du2,Datta}.

In order to state our no-go theorem on  mixed-state conversion. We need to construct a set of {\it new imaginarity measures} using Theorem 1. Let
$$f_{k}(x)=\frac{1-x }{k}\wedge 1,\ \ \forall k\in(0,1],$$ here
$a\wedge 1$ denotes the minimal value of
$a$ and $1$. It is clear that each $f_{k}$
fulfills the conditions (i) and (ii). Since the following equation
$$(\lambda a+(1-\lambda)b)\wedge 1\geq \lambda(a\wedge 1)+(1-\lambda)(b\wedge 1)$$
holds for any $a,b\geq 0$ and any $\lambda\in[0,1]$, one can show
that $f_{k}$ is concave, that is, $f_{ k}$ satisfies the
condition (iii) for any
$k\in(0,1]$. So, by Theorem 1, each ${\mathcal I}_{f_k}$ is also a
imaginarity measure.

Now, we are in a position to give no-go theorem on  mixed-state conversion.

{\bf Theorem 2.} {\it For any finite number of imaginarity measures
$\{	{\mathcal I}_j\}_{j=1}^{s}$ $ (s <\infty)$ satisfying (I1-I4) defined on $d$-dimensional
states, and $d\geq  4$, there exist mixed states $\rho$, $\sigma$ and $k$ such that ${\mathcal I}_j(\rho)\geq{\mathcal I}_j(\sigma)$ $( j = 1,\ldots, s)$ and ${\mathcal I}_{f_k}(\rho)<{\mathcal I}_{f_k}(\sigma )$, i.e., there is no real operation that
converts $\rho$ into $\sigma$.}

From Theorem 2, we see that characterization of
 conversion on general mixed states is complicated
since it involves an infinite number of conditions. In \cite{Datta}, it is shown that there does not exist a finite complete
set of continuous and faithful resource monotones to determine state transformations on general quantum resource theory.
However, by Theorem 2,  even we drop continuous and faithful requirements of imaginarity measures, there still does not exist a finite complete
set to determine mixed-state conversion on resource theory of imaginarity.
Thus we need to find more powerful imaginarity measures with operational meaning and suitable free operations in order to classify mixed-state conversion \cite{Regula1,Regula2,Du4}.

In the following, we are devoted to presenting another imaginarity monotone in terms of a state conversion process.
 We consider that some pure states undergo real operations
 and finally become the common objective state. It will be
shown that the imaginarity of the objective state can be well described by the least imaginarity of the pure input states.

By \cite{Kondra}, any state can be produced from some input pure states via the
corresponding real operations.  Denote ${\mathfrak R}(\rho)$ is the set of pure states that can
be converted into the given state $\rho$ by RO. For every function $f$ with (i)-(iii), we define \begin{equation}\widetilde{\mathcal I}_f(\rho) = \min_
	{|\phi\ra \in {\mathfrak R}(\rho)}\mathcal I_f (|\phi\ra).
\end{equation}

$\widetilde{\mathcal I}_f(\rho)$ can be understood as imaginarity cost, the formation process we
prepares a mixed state by consuming pure imaginarity states
under real operations. It is originated from coherence cost in the one-shot and asymptotic limit case \cite{Winter, Zhao,Takagi}. The
coherence cost happened to be equal to the coherence of
formation, a single-letter formula, which is equivalent to
the convex roof construction based on the relative entropy
coherence of a pure state \cite{Winter}. In this sense, the coherence cost can be regarded as specific coherence measure (relative entropy) for pure states.
However, imaginarity monotone in the current paper essentially is given by
the least imaginarity of the pure state that can be converted to
the given state by real operations. Additionally, The establishment
of Eq. (13) strongly depends on the imaginarity measure
for pure states.

{\bf Theorem 3.}  {\it The $\widetilde{\mathcal I}_f(\rho)$ satisfies {\rm(I1)-(I3)}.}

By Theorem 3, a valid imaginarity monotone $\widetilde{\mathcal I}_f(\cdot)$ has been
completely established if $f$
is given. Moreover, one can see that
$\widetilde{\mathcal I}_f(\rho)$  can be obtained by the minimal pure-state
imaginarity optimized in the set $R(\rho )$. Furthermore, $\widetilde{\mathcal I}_f(|\psi\rangle)={\mathcal I}_f(|\psi\rangle)$
and $\widetilde{\mathcal I}_f(\rho)\geq{\mathcal I}_f(\rho)$.
In fact, given
any imaginarity monotone  ${\mathcal I}$ defined on pure states,
the imaginarity monotone extended to mixed states through
our method serves as the supremum of all the imaginarity
monotones equal to ${\mathcal I}$ for pure states.  A parallel result in coherence theory can be found in \cite{Yu}.
If we select the geometric imaginarity measure \cite{Guo1,Guo2,Kondra} to quantify the imaginarity
of pure states,  then Theorem 3  endows
an operational understanding to the geometric imaginarity.

By the definition of $\widetilde{\mathcal I}_f(\rho)$, in fact, $\widetilde{\mathcal I}_f(\rho)$
can be expressed in in terms of pure-state decompositions of $\rho$.

{\bf Theorem 4.} {\it $\widetilde{\mathcal I}_f(\rho) = \min_{\{p_i, |\phi_i\ra\}}f (\sum_i p_i|\la\phi^*_i|\phi_i\rangle|),$ the min runs over all  pure-state decomposition of $\rho$}

From Theorem 4, it is evident  \begin{equation}\widetilde{\mathcal I}_f(\rho) =f(\max_ {\{p_i, |\phi_i\ra\}}\sum_i p_i|\la\phi^*_i|\phi_i\rangle|) .\end{equation}
Moreover, one can find that to obtain the analytic expression of $\widetilde{\mathcal I}_f(\rho)$, whether there
exists the optimal pure-state decomposition of $\rho$ is key.
In the next, we find the optimal pure-state decomposition of $\rho$ for the qubit case,
and so give the analytic formula of $\widetilde{\mathcal I}_f(\rho)$.

{\bf Theorem 5.} {\it Given an imaginary state  $\rho$ of a qubit with $b=e^{\rm i\theta}|b|$
denoting its off-diagonal element, the optimal decomposition
can be given by
\begin{equation}\rho=\lambda |\psi_+\ra\la\psi_+|+(1-\lambda)|\psi_-\ra\la\psi_-|,\end{equation}
where $|\psi_{\pm}\ra=e^{\rm i\theta}\sqrt{\frac{1\pm z}2 }|1\ra +{\rm i}\sqrt{\frac{1\mp z}2}|2\ra$ with $z=\sqrt{1-4|b|^2}$, and $\lambda\in [0,1]$
is some weight parameter determined by the state $\rho$. In this sense, \begin{equation}\widetilde{\mathcal I}_f(\rho) =f(\sqrt{1-4({\rm Im} b)^2}),\end{equation}
${\rm Im} b$ is the imaginary part of $b$.}

Now we turn to the convexity of $\widetilde{\mathcal I}_f$ which does not hold true generally from Theorem 3.  We will study the requirements of  $f$
such that $\widetilde{\mathcal I}_f$ is convex. In fact, for a state $\rho$, it is easy to see that $\widetilde{\mathcal I}_f(\rho) = \mathcal I_f (\rho)$ is equivalent to that $\widetilde{\mathcal I}_f$ is convex by the definition of $\widetilde{\mathcal I}_f$. From the observation,  we get the following result.

{\bf Theorem 6.} {\it The following statements are equivalent:\\
  (1) $\widetilde{\mathcal I}_f$ is convex;\\
   (2) $\widetilde{\mathcal I}_f(\rho) = \mathcal I_f (\rho) $;\\
   (3) There always exists
   a pure state ensemble $\{\tilde{p}_i, |\tilde{\psi}_i\ra\}$ such that $f(\sum_i \tilde{p}_i |\la \tilde{\psi}_i^*|\tilde{\psi}_i\ra|))\leq \sum_i p_i f(|\la \psi_i^*|\psi_i\ra|)$,
   here  $\{p_i, |\psi_i\ra\}$ is any ensemble of $\rho$.}

By Theorem 6,  if the condition given by (3) are satisfied, then the proposed imaginarity monotone $\widetilde{\mathcal I}_f$ is the
the imaginarity measure ${\mathcal I}_f $ in terms of the convex roof construction. Let $f(x)=\frac{1-x}{2}$,
assume  $\{\tilde{p}_i, |\tilde{\psi}_i\ra\}$ be the  optimal decomposition of $\max_ {\{p_i, |\phi_i\ra\}}\sum_i p_i|\la\phi^*_i|\phi_i\rangle|$,
because  $f$ is monotonous decrease convex  function, therefore (3) is satisfied. By Theorem 1, we have known that ${\mathcal I}_f$ is the geometric imaginarity \cite{Guo1,Guo2,Kondra}.  This tells that the geometric imaginarity can  be presented in terms of a state conversion process. This  provides operational meaning to the geometric imaginarity.

For general $f$,  the condition (3) is a quite demanding  mathematical requirement. we will give a numerical comparison to show  $\widetilde{\mathcal I}_f(\tau)>  {\mathcal I}_f(\tau)$ for some qutrit imaginary states (see Fig. 1).
Given a density matrix $\tau=\lambda |\phi\ra\la\phi|+(1-\lambda)|3\ra\la3|$, where $|\phi\ra= \sqrt{\frac{1+z}2}|1\ra-{\rm i}\sqrt{\frac{1-z}2}|2\ra (0\leq z<1)$,
%with $|i\ra_{ i=1,2,3}$ denoting the computational basis of the 3-dimension Hilbert space ${\mathcal H}$ ,
it will be shown in the appendix (see Proposition 4),  $$\widetilde{\mathcal I}_f(\tau)=f(\lambda z+1-\lambda).$$
where $f$ is any given function satisfying (i-iii).
%Thus we can show the difference between $\widetilde{\mathcal I}_f$ and ${\mathcal I}_f$.
Note that ${\mathcal I}_f$
satisfies (I5), one can get ${\mathcal I}_f(\tau)=\lambda f(z)$, which is usually not equal to $\widetilde{\mathcal I}_f$. To explicitly demonstrate the difference, let us select $f(x)=\sqrt{\frac {1-x^2}2}$. The numerical comparison is plotted in Fig. 1(a), which demonstrates the
apparent difference. In addition, we also consider the imagenarity based on the relative entropy. In this case, $f(x)= -\frac{1+x}2\log (\frac{1+x}2)-\frac{1-x}2\log (\frac{1-x}2) $.  Their numerical comparison is shown
in Fig. 1(b), which also illustrates their difference.
\begin{figure}[htbp]
	\centering
\subfigure{\includegraphics[scale=0.45]{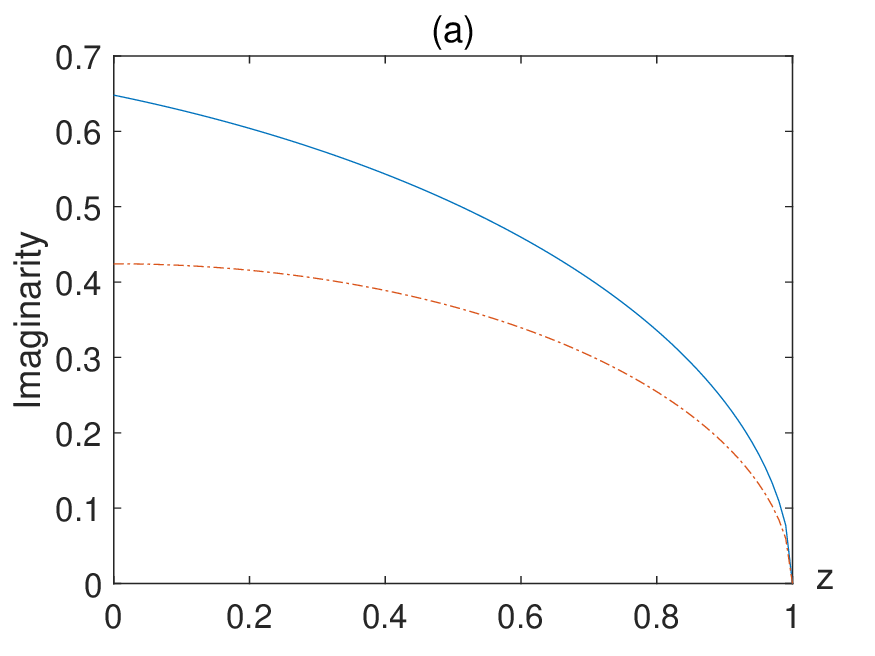}}
 \subfigure{\includegraphics[scale=0.45]{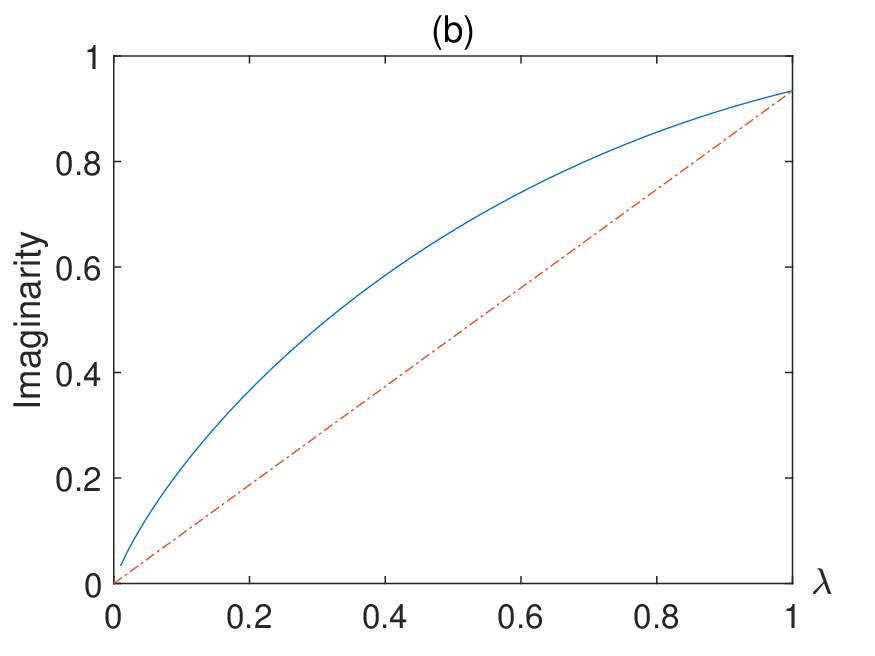}}
		\caption{\small Comparisons between $\widetilde{\mathcal I}_f(\tau)$ (solid line) and ${\mathcal I}_f(\tau)$ (dot-dashed line), (a) $\lambda=0.6, z\in[0,1], (b) z=0.3, \lambda\in[0,1]$. }
\end{figure}

\section {Conclusions and discussions} In this work, we have provided
 two approaches to quantify imaginarity. In the first, we have built the framework of the convex roof construction in resource theory of imaginarity, a traditional and
effective method in the quantification of entanglement
 and coherence. Moreover,  it has been shown the restriction of any imaginarity measure satisfying (I1), (I3), (I4)) to pure states
is identical to some ${\mathcal I}_f $. Based on this method, one can construct many interesting imaginarity measures, such as the geometric imaginarity and so on.
 We also apply  the tool to study the state
conversion problem in resource theory of imaginarity. It is found that there does not exist a finite complete set of imaginarity measures determining mixed-state conversion. Secondly, we present the  imaginarity monotone in terms of a state conversion
process under real operations. In particular, it
is shown that our imaginarity monotone is the supremum
of all imaginarity monotones with the same imaginarity for any given pure state. An interesting observation is that the geometric imaginarity
can be presented  by this approach which provides operational meaning to the geometric imaginarity.
 The inner relation between  quantifiers of
imaginarity by the convex roof construction and quantifiers of the imaginarity by the least imaginarity of the input pure states has also discussed.

A closely related area with quantifying imaginarity is how to detect  imaginarity for a tupe of quantum states $(\rho_1, \rho_2, \ldots, \rho_n)$ (see \cite{Wagner3} and references therein). Our results raise an interesting question  how to quantifying imaginarity of $(\rho_1, \rho_2, \ldots, \rho_n)$? It extends the definition of imaginarity  measure and  may develop new methods for quantifying imaginarity.

\vspace{0.1in}
{\it Acknowledgement.---}
The authors thank referees for providing constructive comments which improve the presentation of the study.
The authors thank professor Jianwei Xu for helpful comments during the preparation of this paper. We acknowledge that this research was supported by NSF of China (12271452), NSF of Xiamen (3502Z202373018), and NSF of Fujian (2023J01028).

\section{APPENDIX: PROOF OF MAIN RESULTS}

 We  need some propositions to give proofs of our theorems.

The proposition 1 is from \cite{Guo2} which provides the generic form of pure states under real unitary operation which is very relevant  to imaginarity
quantification for pure states.

 {\bf Proposition 1.} {\it For any pure state $|\psi\ra$, there exists a real orthogonal matrix $O$ such that
$O|\psi\ra=\sqrt{\frac{1+|\la \psi^*|\psi\ra|}{2}} |1\ra+\rm i\sqrt{\frac{1-|\la \psi^*|\psi\ra|}{2}} |2\ra$.}

The Proposition 2 is the realization of conversion of ensembles under real operations.

{\bf Proposition 2.} {\it For ensemble $\{p_j,|\psi_j\ra\}_{j=1}^m$, there is real operation $\Phi$ with Kraus operators $A_j$ such that $$A_j|\eta\ra=\sqrt{p_j}|\psi_j\ra,\eqno{(A1)}$$ here $|\eta\ra =\sqrt{\frac{1+x}{2}}|1\ra +{\rm i}\sqrt{\frac{1-x}{2}}|2\ra$, $x=\sum_{j=1}^mp_j|\la \psi_j^*|\psi_j\rangle|$.}

{\bf Proof.}  Write
$x_j=|\la \psi_j|\psi_j\rangle|$,  $|\eta_j\ra =\sqrt{\frac{1+x_j}{2}}|1\ra +{\rm i} \sqrt{\frac{1-x_j}{2}}|2\ra$.
For any $1\leq j\leq m$, define
$$K_j=\sqrt{p_j}\text{diag}(\frac{\sqrt{\frac{1+x_j}{2}}}{\sqrt{\frac{1+x}{2}}},\frac{\sqrt{\frac{1-x_j}{2}}}{\sqrt{\frac{1-x}{2}}},1,\ldots,1).$$
It is easy to check that $\sum_{j=1}^mK_j^\dag K_j=I_d$ and
$K_j$ is a real operator.  In
addition, a simple calculation yields
$$K_j |\eta\rangle=\sqrt{p_j}|\eta_j\rangle.$$
Form Proposition 1, it follows that there exists a real
orthogonal matrix $O_j$ such that $O_j|\eta_j\ra=|\psi_j\ra$. Let $A_j=O_jK_j$, then  the
map $\Phi$ defined by
$\Phi(\cdot)=\sum_{j=1}^m A_j(\cdot )A^{\dag}_j$ is a real operation with $A_j|\eta\ra=\sqrt{p_j}|\psi_j\ra$.

{\bf Proposition 3.} {\it If we write a pure state $\rho=|\psi\ra\la \psi |=\sum_{m,n} \rho_{mn}|m\ra\la n|$, here $\rho_{mn}=\la m|\rho |n\ra$, $\{|m\ra\}_{m=1}^d$ is the prefixed basis, then $$\label{eq1}|\la\psi^*|\psi\ra|=\sqrt{1-2\sum_{m\neq n}(\rm{Im}\rho_{mn})^2}.\eqno(A2)$$

{\bf Proof.} Let $|\psi\ra=\sum_m (a_m+{\rm i}b_m)|m\ra$ with $\sum_m (a_m^2+b_m^2)=1$ and $a_m,b_m\in \bb R$.  Then
$$\begin{array}{ll}
&|\la\psi^*|\psi\ra|^2\\
=&|\sum_m (a_m+{\rm i}b_m)^2|\\
=&(\sum_m a_m^2-b_m^2)^2+4(\sum_m a_mb_m)^2,\end{array}$$
$$\sum_{m\neq n}(\rm{Im}\rho_{mn})^2=\sum_{m\neq n} (a_nb_m-a_mb_n)^2.$$
A direct computation shows that
$$|\la\psi^*|\psi\ra|^2+2\sum_{m\neq n}(\rm{Im}\rho_{mn})^2=(\sum_m (a_m^2+b_m^2))^2=1.$$

{\bf Proposition 4.} {\it Given a density matrix $\tau=\lambda |\phi\ra\la\phi|+(1-\lambda)|3\ra\la3|$, where $|\phi\ra= \sqrt{\frac{1+z}2}|1\ra-{\rm i}\sqrt{\frac{1-z}2}|2\ra (0\leq z<1)$, with
$|i\ra_{ i=1,2,3}$ denoting the computational basis of the 3-dimension,
it is shown that $$\widetilde{\mathcal I}_f(\tau)=f(\lambda z+1-\lambda).$$
where $f$ is any given function satisfying (i-iii).}}

{\bf Proof.} From decreasing property of $f$, in order to finish the proof, we only need to verify that, for every pure-state decomposition $\{p_i, |\phi_i\ra\}$ of $\tau$,
$$\sum_i p_i|\la\phi^*_i|\phi_i\rangle|\leq \lambda z+1-\lambda.$$ Write  $|\phi_i\ra=x_i|1\ra+y_i|2\ra+z_i|3\ra$.
In particular, the parameters $x_i$, $y_i$, $z_i$
satisfy
$$\label{eq4}
	\left\{ \begin{array}{l}
\sum_i p_i |x_i|^2=\frac{\lambda(1+z)}2,\\	
\sum_i p_i |y_i|^2=\frac{\lambda(1-z)}2,\\	
\sum_i p_i |z_i|^2=1-\lambda,\\	
\sum_i p_i x_iy_i^*={\rm i}\frac{\lambda\sqrt{1-z^2}}2.	
	\end{array} \right.  (A3)
$$
It shows that $$|\sum_i p_i x_iy_i^*|^2=(\sum_i p_i |x_i|^2)(\sum_i p_i |y_i|^2).$$ The saturation of the Cauchy-Schwarz inequality  requires $x_i =
gy_i$ for any $i$, with $g$ being a nonzero constant. Thus it is not
difficult to find
$g={\rm i}\sqrt{\frac{1+z}{1-z}}$ from (A3).
So \begin{widetext} $$\begin{array}{ll}
	&\sum_i p_i|\la\phi^*_i|\phi_i\rangle| \\
	= &\sum_i p_i\sqrt{1-4((\text{ Im}(x_iy_i^*))^2+(\text{ Im}(x_iz_i^*))^2+(\text{ Im}(y_iz_i^*))^2)}\\
= &\sum_i p_i\sqrt{1-4(|g|^2|y_i|^4+|g|^2(\text{ Re}(y_iz_i^*))^2+(\text{ Im}(y_iz_i^*))^2)}\\
\leq & \sum_i p_i\sqrt{1-4(|g|^2|y_i|^4+|y_i|^2|z_i|^2)}\\
=& \sum_i p_i\sqrt{(|x_i|^2+|y_i|^2+|z_i|^2)^2-4(|g|^2|y_i|^4+|y_i|^2|z_i|^2)}\\
=& \sum_i p_i\sqrt{((|g|^2+1)|y_i|^2+|z_i|^2)^2-4(|g|^2|y_i|^4+|y_i|^2|z_i|^2)}\\
=& \sum_i p_i\sqrt{((|g|^2-1)|y_i|^2+|z_i|^2)^2}\\
=&\sum_i p_i((|g|^2-1)|y_i|^2+|z_i|^2)\\
=& \lambda z+1-\lambda,
\end{array}$$\end{widetext}
where the first equality is  from (A2).

Now we are in a position to give proofs of our theorems

{\bf Proof of Theorem 1.}	 (I1: Nonnegativity)  If $\rho$ is  an arbitrary real state, then there is
a real decomposition $\rho=\sum_i p_i|\psi_i\ra\la \psi_i|$ such that ${\mathcal I}_f(|\psi_i\rangle)=0$.
By the definition of ${\mathcal I}_f(\rho)$, we have ${\mathcal I}_f(\rho)=0$.

(I3: Probabilistic monotonicity) We first prove it for
pure states. Note that, for real operation $\Phi$ with Kraus operators $K_n$, $\sum_n|\la  \psi^*K_n^T|K_n\psi\ra|\geq  |\la\psi^*|\psi\ra|$.
Because $f$ is decreasing and concave, it follows that
$$\begin{array}{ll}
	& {\mathcal I}_f(|\psi\rangle)=f(|\la \psi^*|\psi\ra|)\\
	\geq &f(\sum_n|\la  \psi^*|K_n^TK_n|\psi\ra|)\\
	\geq & \sum_n {\text Tr}(K_n |\psi\ra\la \psi| K_n^{\dag}) f(\frac{|\la  \psi^*| K_n^TK_n|\psi\ra |}{ {\text Tr}(K_n |\psi\ra\la \psi| K_n^{\dag})})\\
	= &  \sum_n {\text Tr}(K_n |\psi\ra\la \psi| K_n^{\dag}) \mathcal I_f(\frac{K_n |\psi\ra\la \psi| K_n^{\dag}}{ {\text Tr}( K_n |\psi\ra\la \psi| K_n^{\dag})} ) .
\end{array}$$

To extend this result to mixed states, consider an
optimal decomposition of a mixed state $\rho=\sum_jp_j |\psi_j\ra\la \psi_j|$,
such that $$ {\mathcal I}_f(\rho)=\sum_jp_j {\mathcal I}_f(|\psi_j\ra\la \psi_j|).$$
Introducing the quantity $c_{nj} =\la\psi_ j |K_n^TK_n|\psi_j\ra$ and $q_n=\text{  Tr}(K_n \rho K_n^{\dag})$, we obtain
$$\begin{array}{ll}
	&\sum_n q_n {\mathcal I}_f(\frac{K_n \rho K_n^{\dag}}{q_n})\\
	=& \sum_n q_n {\mathcal I}_f(\sum_j p_j \frac{K_n |\psi_j\ra\la \psi_j| K_n^{\dag}}{q_n})\\
	=& \sum_n q_n {\mathcal I}_f(\sum_j \frac {p_jc_{nj}}{q_n}\times  \frac{K_n |\psi_j\ra\la \psi_j| K_n^{\dag}}{c_{nj}})\\
	\leq & \sum_{nj} p_jc_{nj} {\mathcal I}_f(\frac {K_n |\psi_j\ra\la \psi_j| K_n^{\dag}}{c_{nj}})\\
	\leq & \sum_j p_j\mathcal I_f(|\psi_j\ra) =\mathcal I_f(\rho).\end{array}$$

(I4: Convexity)  Let $\rho=\sum_i p_i\rho_i$ be any
ensemble of $\rho$. And let $\rho_i=\sum_j q_{ij}\rho_{ij}$ be an
optimal pure-state ensemble of $\rho_i$, i.e., ${\mathcal I}_f(\rho_i)=\sum_j
q_{ij}{\mathcal I}_f(\rho_{ij})$. Then
$$\begin{array}{ll}
		& {\mathcal I}_f(\rho)={\mathcal I}_f(\sum_ip_i\sum_jq_{ij}\rho_{ij})\\
		=& {\mathcal I}_f(\sum_{ij}p_iq_{ij}\rho_{ij})\leq
		\sum_{ij}p_iq_{ij}{\mathcal I}_f(\rho_{ij})\\
		= & \sum_i p_i {\mathcal I}_f(\rho_i).\end{array}$$ The
inequality follows from the definition of ${\mathcal I}_f$.

Let $\mu$ be an arbitrary imaginarity measure.
Define $f$ by
$f(x)=\mu(|\psi\rangle\langle \psi|)$, where
$x=|\la \psi^*|\psi\ra|$. Note that if $\la \psi^*|\psi\ra|= |\la \phi^*|\phi\ra|$ then $\mu(|\psi\ra) =\mu(|\phi\ra)$, we get the $f$ is well-defined.
In the following, we will check that $f$ satisfies
conditions (i)-(iii). The (i) and  (ii) follows from (I1) and (I2) respectively.
To prove
(iii), for $x,y\in [0,1]$, let
$$|\psi_1\ra=\sqrt{\frac{1+x}{2}}|1\ra+{\rm i}\sqrt{\frac{1-x}{2}}|2\ra,$$
$$|\psi_2\ra =\sqrt{\frac{1+y}{2}}|1\ra+{\rm i}\sqrt{\frac{1-y}{2}}|2\ra,$$
$$|\phi\ra=\sqrt{\frac{1+\lambda x+(1-\lambda)y}{2}}|1\ra+{\rm i}\sqrt{\frac{1-(\lambda x+(1-\lambda)y)}{2}}|2\ra,$$ here $\lambda\in [0,1]$.
It is easy to check that $|\la \psi_1^*|\psi_1\ra|=x$, $|\la \psi_2^*|\psi_2\ra|=y$ and $|\la \phi^*|\phi\ra|=\lambda x+(1-\lambda)y$.
Define
{\scriptsize $$ K_1=\sqrt{\lambda} \text{ diag}(\frac{\sqrt{\frac{1+x}{2}}}{\sqrt{\frac{1+\lambda x+(1-\lambda)y}{2}}}, \frac{ \sqrt{\frac{1-x}{2}}}{\sqrt{\frac{1-(\lambda x+(1-\lambda)y)}{2}}},1, \cdots,1),$$
	$$K_2=\sqrt{1-\lambda} \text{ diag}(\frac{\sqrt{\frac{1+y}{2}}}{\sqrt{\frac{1+\lambda x+(1-\lambda)y}{2}}}, \frac{ \sqrt{\frac{1-y}{2}}}{\sqrt{\frac{1-(\lambda x+(1-\lambda)y)}{2}}},1, \cdots,1).$$ }
It is easy to check that
$\Phi(\cdot)=K_1\cdot K_1^{\dag}+K_2\cdot K_2^{\dag}$ is a real
operation. And
$$K_1|\phi\ra=\sqrt{\lambda}|\psi_1\ra,\ \ \ K_2|\phi\ra=\sqrt{1-\lambda}|\psi_2\ra.\ \ \  (A4) $$
From (I3 Probabilistic monotonicity), we get that
$$
	\lambda \mu(|\psi_1\ra)+(1-\lambda) \mu( |
	\psi_2\rangle)
	\leq   \mu
	(|\phi\ra). \ \ \ \ \ \ \ \  \ (A5)$$ That is $f(\lambda
{ x}+(1-\lambda ){ y})\geq \lambda f({ x})+(1-\lambda)f({
	y})$, i.e., $f$ is concave.

{\bf Proof of Theorem 2.} Take $\rho=p_1|\psi_1\ra\la\psi_1|+p_2|\psi_2\ra\la\psi_2|$, where $p_1$ and $p_2$ will be determined later, and    $|\psi_1\ra=\frac 1{\sqrt 2}(|1\ra+{\rm i}|2\ra)$ and $|\psi_2\ra=\sqrt{\lambda}|3\ra+\sqrt{1-\lambda}{\rm i}|4\ra$, where the $\lambda$ is less than $\frac 12$ and will be determined later. The density matrix $\sigma$ is taken to be pure state: $\sigma_{\eta}=|\phi_{\eta}\ra\la\phi_{\eta}|$, here $|\phi_{\eta}\ra=\sqrt{\eta}|1\ra+\sqrt{1-\eta}{\rm i}|2\ra$ ($\eta<\frac 12$). Thus, from (I5), any measure of imaginarity ${\mathcal I}_j$ must satisfy ${\mathcal I}_j(\rho)= p_1 {\mathcal I}_j(|\psi_1\ra\la\psi_1|)+p_2 {\mathcal I}_j(|\psi_2\ra\la\psi_2|)$.

We can suppose, for $j=1,2,\cdots,s_1$, ${\mathcal I}_j(|\psi_1\ra\la\psi_1|)> {\mathcal I}_j(|\phi_{\eta_j}\ra\la\phi_{\eta_j}|)$ for some $\eta_j$ and, for other $j$, ${\mathcal I}_j(|\psi_1\ra\la\psi_1|)= {\mathcal I}_j(|\phi_{\eta}\ra\la\phi_{\eta}|)$ for all $\eta\neq 0$. Since the number of ${\mathcal I}_j$ is finite, there exists $p_1$ such that $1>p_ 1>\frac{ {\mathcal I}_j(|\phi_{\eta_0}\ra\la\phi_{\eta_0}|)} {{\mathcal I}_j(|\psi_1\ra\la\psi_1|)}$, for $j=1,2,\cdots,s_1$, here  $\eta_0=\min_{1\leq j\leq s_1}\{\eta_j\}$. Thus ${\mathcal I}_j(\rho)= p_1 {\mathcal I}_j(|\psi_1\ra\la\psi_1|)+p_2 {\mathcal I}_j(|\psi_2\ra\la\psi_2|)\geq {\mathcal I}_j(\sigma_{\eta_0})$ for all $j=1,2,\cdots, s_1$. Clearly, this holds true for $j=s_1+1,\cdots ,s$.

On the other hand, if  there is a real operation $\Phi$ such that $\Phi(\rho)=\sigma_{\eta_0}$, then ${\mathcal I}_{f_{k}}(\rho)\geq {\mathcal I}_{f_{k}}(\sigma_{\eta_0})$.
That is $$p_1+p_2(\frac{1-|\la \psi_2^*|\psi_2\ra| }{k}\wedge 1)\geq\frac{1-|\la\phi_{\eta_0}^*|\phi_{\eta_0}\ra| }{k}\wedge 1.$$
Choosing  $k=1-|\la \phi_{\eta_0}^*|\phi_{\eta_0}\ra|$, one can get $|\la\psi_2^*|\psi_2\ra|\leq |\la\phi_{\eta_0}^*|\phi_{\eta_0}\ra| $. But for $\lambda >\eta_0$, that is not true.

{\bf Proof of Theorem 3.} (I1: Nonnegativity)  If $\sigma$ is  an arbitrary real state, then there is a
a real decomposition $\rho=\sum_i p_i|\psi_i\ra\la \psi_i|$. From Proposition 2,  there is real pure state $|\psi\ra $ and real operation $\Phi$  such that
$\Phi(|\psi\ra)=\rho$.
So, $\widetilde{\mathcal I}_f(\rho)=0$.

(I2: Imaginarity monotonicity)
Let $\Phi$ be an arbitrary RO and $\rho$ denote
any state. Suppose $|\psi\ra\in {\mathfrak R}(\rho)$ is the optimal pure state subject
to $\widetilde{\mathcal I}_f (\rho) =\widetilde{\mathcal I}_f(|\psi\ra)$, then it implies that $\Phi_0(|\psi\ra\la\psi|)=\rho$
for some real operation $\Phi_0$. Then $\Phi(\rho)=\Phi(\Phi_0(|\psi\ra))$. By
definition of $\widetilde{\mathcal I}_f$, one can easily find $\widetilde{\mathcal I}_f(\rho)=\widetilde{\mathcal I}_f(|\psi\ra)\geq\widetilde{\mathcal I}_f(\Phi(\rho)$.

(I3: Probabilistic monotonicity)  Since $\widetilde{\mathcal I}_f(|\psi\ra)=\mathcal I_f(|\psi\ra)$ for all pure state $|\psi\ra$, probabilistic imaginarity monotonicity is true  for
pure states.

To extend this result to mixed states, consider a mixed state $\rho$ and a real operation $\Phi(\cdot)=\sum_n K_n\cdot K_n^{\dag}$.
Let $$p_n=\text{Tr}K_n\rho K_n^{\dag},$$ $$\rho_n=\frac{K_n\rho K_n^{\dag}}{p_n},$$ and
$|\psi\ra$ be an optimal state in ${\mathfrak R}(\rho)$
such that $\widetilde{ {\mathcal I}}_f(\rho)={\mathcal I}_f(|\psi\ra)$,  $\rho=\sum_jA_j|\psi\ra\la\psi|A_j^{\dag}$.
Introducing the quantity $c_{nj} =\la\psi |K_n^TA^{T}_ jA_jK_n|\psi\ra$, we obtain
$$\begin{array}{ll}
	&\sum_n p_n \widetilde{ {\mathcal I}}_f(\frac{K_n \rho K_n^{\dag}}{p_n})\\
	=& \sum_n p_n \widetilde{ {\mathcal I}}_f(\sum_j  \frac{K_nA_j |\psi\ra\la \psi|A_j^{\dag} K_n^{\dag}}{p_n})\\
	=& \sum_n p_n \widetilde{{\mathcal I}}_f(\sum_j \frac {c_{nj}}{p_n}\times  \frac{K_nA_j |\psi\ra\la \psi| A_j^{\dag}K_n^{\dag}}{c_{nj}})\\
	\leq & \sum_{n} p_nf(\frac {\sum _j |\la K_n^*A_j^* \psi^*|\psi A_jK_n\ra| }{p_{n}})\\
	\leq & f(\sum_{nj}   |\la K_n^*A_j^* \psi^*|\psi A_jK_n\ra| )\\
	\leq& f(|\la \psi^*|\psi\ra|)=\widetilde{\mathcal I}_f(|\psi\ra),\end{array}$$
where the first inequality is from the definition of $\widetilde{ {\mathcal I}}_f$ and Proposition 2, the second inequality is because of the concavity  of $f$, the third comes from Proposition 1 and decreasing property of $f$.

{\bf Proof of Theorem 4.} For each pure ensemble $\{p_i, |\phi_i\ra\}$, using Proposition 2, we obtain that
$$\widetilde{\mathcal I}_f(\rho) \leq \min_{\{p_i, |\phi_i\ra\}}f (\sum_i p_i|\la\phi^*_i|\phi_i\rangle|).$$
On  the other hand, if $\rho=\Phi(|\psi\ra\la \psi|)=\sum_i K_i|\psi\ra\la\psi|K_i^{\dag}$ and $\widetilde{\mathcal I}_f(\rho) = f( |\la\psi^*|\psi\ra|),$ then there exists a decomposition $\{p_i,|\psi_i\ra\}$ with $p_i=\text{Tr} (K_i|\psi\ra\la\psi|K_i^{\dag})$, $|\psi_i\ra\la \psi_i|= \frac{K_i|\psi\ra\la\psi|K_i^{\dag}}{p_i}$. From Note that, $$\sum_i p_i |\la \psi^*_i|\psi_i\ra|\geq |\la\psi^*|\psi\ra|.$$  Then
$$\begin{array}{ll}
&\min_{\{p_i, |\phi_i\ra\}}f (\sum_i p_i|\la\phi^*_i|\phi_i\rangle|)\\
\leq & f(\sum_i p_i |\la \psi^*_i|\psi_i\ra|)\\\leq &f(|\la\psi^*|\psi\ra|)
= \widetilde{\mathcal I}_f(\rho).\end{array}$$

{\bf Proof of Theorem 5.} We only need to show that $$\sum_i p_i|\la\phi^*_i|\phi_i\rangle|\leq \lambda |\la\psi_+^*|\psi_+\ra|+(1-\lambda)|\la\psi_-^*|\psi_-\ra|,$$ for every decomposition $\{p_i, |\phi_i\ra\}$. Similar to the state $|\psi^{\pm}\ra$, we can use $\{b_i,z_i\}$ to
express the states $|\phi_i\ra$ with $z_i =\sqrt{1-4|b_i|^2}$. By a  direct computation, one can obtain that $$|\la\phi^*_i|\phi_i\rangle|=\sqrt{1-4(\text{Im}b_i)^2},$$
$$|\la\psi_{\pm}^*|\psi_{\pm}\rangle|=\sqrt{1-4(\text{Im}b)^2}.$$
$$\begin{array}{ll}
 &\sum_i p_i|\la\phi^*_i|\phi_i\rangle|= \sum_i p_i \sqrt{1-4(\text{Im}b_i)^2}\\
 \leq & \sqrt{1-4(\sum_i p_i\text{Im}b_i)^2}=  \sqrt{1-4(\text{Im}b)^2}\\
 =& \lambda |\la\psi_+|\psi_+\ra|+(1-\lambda)|\la\psi_-|\psi_-\ra|,
\end{array}$$
where the inequality is from function $\sqrt{1-4x^2}$ is concave. From Theorem 4, $\widetilde{\mathcal I}_f(\rho) =f(|\la\psi_{\pm}^*|\phi_{\pm}\rangle|) =f(\sqrt{1-4(\text{Im}b)^2})$.

\end{document}